\documentclass[10pt]{article} 

\def\mee{m_{\rm ee}}

\usepackage{geometry}                		
\usepackage[artemisia]{textgreek}
\usepackage{graphicx}
\usepackage{multicol}        
\usepackage[unicode]{hyperref}
\usepackage{amssymb}
\usepackage[T1]{fontenc}
\usepackage[utf8]{inputenc}
\usepackage{amsmath}
\usepackage{array}
\usepackage{graphics}
\usepackage{graphicx}
\usepackage{array}
\usepackage{fancyhdr}
\usepackage{enumerate}
\usepackage{verbatim}

\usepackage{bm}
\usepackage{amssymb,upgreek}
\usepackage{mathptmx}
\usepackage{multicol,xcolor}

\usepackage{authblk}
\usepackage[numbers,sort&compress]{natbib}

\begin{document}

\title{What is matter according to particle physics and why try to observe its creation in lab} 

\author{\Large Francesco Vissani\\
{\small  \vskip-3ex INFN, Laboratori Nazionali del Gran Sasso, 67100  L'Aquila, Italy {\em and}}\\
{\small \vskip-0.9ex GSSI - Gran Sasso Science Institute, 67100 L'Aquila, Italy}}

\date{}

\maketitle

 \begin{abstract}
\sf  \vskip-10mm 
\noindent The standard model of elementary interactions has long qualified as a theory of matter, in which the postulated conservation laws (one baryonic and three leptonic) acquire theoretical meaning. However, recent observations of lepton number violations - neutrino oscillations - demonstrate its incompleteness.
We discuss why these considerations suggest the correctness of Ettore Majorana's ideas on the nature of neutrino mass, and add further interest to the search for an ultra-rare nuclear process in which two particles of matter (electrons) are created, commonly called neutrinoless double beta decay. The approach of the discussion is mainly historical and its character is introductory. 
Some technical considerations, which highlight the usefulness of Majorana's representation of gamma matrices, are presented in the appendix.
\end{abstract}








\setcounter{secnumdepth}{4}




{

{\footnotesize \tableofcontents}

\parskip0.24ex

\section{Introduction}
The discussion of the nature of things is very old. We can experience the wonder of nature but it is almost inevitable to reason about how to reconcile the evident mutability of what is around us with an equally evident degree of permanence. The hypothesis advanced by Greek atomism is that of a simple and permanent substance, even if not directly perceptible to the senses, which is capable of various arrangements and combinations, that give rise to complex dynamics. It is undeniable that, first modern chemistry and then physics have made dramatic contributions to this discussion, to the point of changing the world in which we live. 

Here we shall examine some speculative questions  raised by modern particle physics pointing in the same direction. Although they do not have the same practical relevance as those of the `science of the atom' properly said, they still deserve some attention, and not only for speculative reasons. For instance, experimental investigations of these questions have not yet led to clear answers and remain a priority to progress, but they are also becoming significantly more challenging, so it makes sense to assess their interest as accurately as possible. Moreover, as just mentioned, this discussion is part of a glorious tradition.

We will reason about 1) what particle physics claims about the nature of matter; 2) what conceptual frameworks it gives us to order the available observations; 3) which ones are the most credible, highlighting those that suggest that matter is to some extent, impermanent. We can already formulate a very precise question to guide the discussion:
 \begin{quote}
{\sf $\star$ Is it possible to observe the creation of matter particles in the laboratory?}
 \end{quote}
It is also interesting to keep in mind a somewhat related question: what is the relationship between the hypothetical processes where particles of matter are created and the equally hypothetical processes in which matter is destroyed. In the following discussion, we will focus mostly on the first question.

\section{Matter and antimatter in particle physics}

\subsection{General features}
Particle physics inherits directly from atomism the idea that sensible reality can be ideally subdivided into well-defined parts, called (elementary) {\em particles}, each with its own characteristics.\footnote{The Greek word {\tt atomos} $(\acute{\alpha}\tau\omicron\mu\omicron\zeta)$
literally means 
`indivisible', the Latin word {\tt particula} means `small part': both concepts refer to atomism. 
The modern usage   
has reserved the first word for the basic entities of chemistry and the second for those of fundamental physics.} 
In the current version, called `relativistic quantum mechanics', particles are described by irreducible representations of finite dimension of the Poincar\'e group, having certain values of spin and mass, and moreover, each particle is accompanied by an antiparticle (with the same mass and opposite charge). This is true for all particles; let us elaborate on the point of antiparticles, since it is very important for the following.

There are various arguments that show how the combination of relativity and wave mechanics results in the necessity of 
{\em antiparticles.} A very attractive one is due to Feynman~\cite{fey} while the one we present is closer to Stuecklberg's \cite{stue}. 
It is not particularly elegant, 
but it has the merit of going straight to the point, clearly highlighting both the undulatory and relativistic bases of the argument. 

\subsection{Waves, relativity and charge} Let us begin from a particle without spin, described by the wave equation of Klein-Gordon
\begin{equation} \label{eq1}
P^2 {\varphi}= m^2 {\varphi}\mbox{ where } P_\upmu=i\hbar \frac{\partial}{\partial x^\upmu}
\end{equation}
In addition to the waves 
with positive energy ${\varphi}_+\sim e^{ -i E t/\hbar}$   there are those with negative energy
 $\sim e^{  +i E t/\hbar}$, that at first sight could seem problematic. 
  
  However, they can be usefully interpreted as follows: if we think of them as conjugate waves $(\varphi_-)^*$, 
 rather than as a wave of ``negative energy'' $\varphi_-$ as one would instinctively do,
we can treat them as outgoing waves -- final states of a transition rather than initial states.\footnote{A transition dipole evolves in time as $x_{1\to 2}\sim e^{ i( E_2-E_1) t/\hbar}$ (Heisenberg); introducing waves with given energy $\psi\sim e^{-iE t/\hbar}$  (de Broglie) it factorizes in a scalar product between two states  
$x_{1\to 2}=\int dx\,  \psi_2^*\,  x\, \psi_1   $ (Schr\"odinger).
Thus the interpretation of the conjugate wave as a final state  arises at basic level.} Since  the electric charge is included according to the `minimal replacement' principle of Weyl, Fock, etc.\ 
(see \cite{oki} for a review) i.e., 
\begin{equation} \label{eq2}
 P_\upmu\to P_\upmu- q\, A_\upmu
\end{equation}
and  since the electromagnetic 
4-potential  $A_\upmu$ is real, the wave ${\varphi}_-$ obeys the equation with $q\to -q$, namely, it has opposite 
charge: it is a different particle, with the same mass, but with oppositie charge. It is an antiparticle. (We show the factor $\hbar$ explicitly in this section, but will use the system $\hbar=c=1$ elsewhere in these notes).

The very same observation applies to Dirac equation, that is
\begin{equation} \label{eq3} ( \mbox{\textgamma}_\upmu\, P^\upmu-m ) {\psi}=0
\end{equation}
The argument is simplified by the Majorana representation of gamma matrices \cite{ettore} (see also appendix~\ref{maj}),  
for which the gamma matrices are purely imaginary, namely 
\begin{equation}\mbox{\textgamma}_\upmu^*=-\mbox{\textgamma}_\upmu\end{equation} 
that we will always use  in these notes.
In fact, with this choice we can interpret the 
``negative energy solutions''  just as above 
\begin{equation}{\psi}_- ^*   \sim  e^{ +i E t/\hbar} \end{equation} 
From
$  [ \mbox{\textgamma}_\upmu (P^\upmu - q \, A^\upmu )  -m ] \, {\psi}_- ^*=0$ we get immediately $[ \mbox{\textgamma}_\upmu (P^\upmu+ q\, A^\upmu ) - m ]\, {\psi}_-=0$;
or, in words, there must be solutions with the same mass and opposite charge (=antiparticles).
The seemingly simple nature of this argument should not mislead the reader.
Even Dirac, who predicted the existence of the anti-electron, needed time to formulate it.\footnote{The point of the paper of 1928 \cite{q928,q928b} 
was to reach a definitive understanding of the spin of the electron. Later, in 1931 he wrote {\sc``A hole, if there were one, would be a new kind of particle, unknown
to experimental physics, having the same mass and opposite charge
to an electron. We may call such a particle an anti-electron''}  \cite{q931};
note in passing the term `hole', that refers to a specific 
interpretation, now largely abandoned.}
An appropriately chosen formalism can help; the conceptual point remains nontrivial.

\subsection{Matter particles}
Let us now focus on those particles that constitute {\em matter}. Their prototype is the electron, the first of them to be discovered and presumably the most important in practice, being the one that makes up the shells of atoms. 

Matter particles:
\begin{itemize}{\sf 
\item[$\star$]  have spin 1/2 and are subject to Fermi-Dirac statistics
\item[$\star$]  are divided into two broad classes: quarks \cite{quark1,quark2}, which have strong interactions and are bound in hadrons; leptons, which do not.}
\end{itemize}
 They also satisfy further conservation laws, 
which we define and discuss next.

In electromagnetic and strong interactions, pairs of particles and antiparticles can be created. This does not imply a violation of the electric charge, and therefore neither a net creation of matter particles. In weak interactions apparently the situation is different. 
It should be stressed that, in the 1930s, Fermi's theory was regarded as revolutionary  \cite{yang} as it led to think about the appearance/disappearance of matter particles for the first time. 
In fact, the very name beta decay suggests the appearance of an electron in the final state, although this is expressed with terminology from days gone by (that of Rutherford). Let us elaborate on this point, discussing the manner how the naive idea that `each particle be forever' was replaced\footnote{This idea was only slowly abandoned. For example, Pauli himself in 1930 believed that neutrinos were contained in the nucleus before decay.} 
and  upgraded by modern particle physics.

After this discovery and even more in 1940s, with the new world of elementary particles, 
a number of questions were raised:  Why a proton does not disintegrate into an anti-electron and a neutral pion? 
 Why a muon does not transform into an electron and a photon? The fundamental laws of conservation of energy, angular momentum or electric charge are compatible with such processes - but they do not happen. The modern formulation of the answer came in 1949 \cite{wig}, when Wigner proposed to consider the existence of a new law governing the possible transformations of matter, called conservation of baryonic number: the number of heavy particles, such as protons and neutrons,
 stays the same in any reaction.  
 A few years later it was suggested that similar laws apply to leptons \cite{lepi1,lepi2,lepi3}.\footnote{According to TD~Lee \cite{td} Fermi knew this in advance. Indeed, the same principles had already been 
 anticipated  by Weyl~\cite{wf} and described by Stueckelberg \cite{slh} albeit talking of `number of heavy particles' rather than `baryon number'.}
 
For instance, returning to weak interactions, 
let us consider a reference case,  
neutron decay, 
\begin{equation} {\rm n\to p+ e+} \bar{\nu}_{\rm e}\end{equation} 
The number of baryons is 1 in the initial and final state, but also  the number of leptons remains unchanged; 
in fact, the newly created electron counts $+1$, but the antineutrino counts $-1$, so the net lepton number  is 0.
We state these laws by saying that the net number of baryons {\bf B} never changes\footnote{In current understanding, baryons are those bound state
 that contain three valence quarks.   Thus, we can express the same in terms of the net number of quarks; a quark has ${\bf B}=1/3$.}; and in a similar way the net number of leptons {\bf L} is always unchanged. 
\begin{quote}
{\sf $\star$ These laws apply to all known interactions and they are part of the current definition of what matter particles are.} 
\end{quote}
Currently we know of no exceptions to these rules. 
Two remarks are in order
\begin{itemize}
\item Experimental tests of {\bf L} are less easy than those of {\bf B}, 
in particular because of the elusive nature of neutrinos, but they are rather as important, as we will see below.
\item There is an interesting theoretical question, about the origin in the known universe of the excess of baryons with respect to the number of anti-baryons.  
\end{itemize}
It is worth noting that both remarks point to major unresolved issues in particle physics; we will come back on that later, 
showing that  {\bf B}  and  {\bf L} are much more closely related than it might seem from these empirical considerations. 

%
%

\section{Are neutrinos particles of matter or are they not?}
\subsection{Majorana's hypothesis} Let us consider again Eqs.~\ref{eq1}-\ref{eq2}. If the particle has no charge, $q=0$, one can identify
the waves ${\varphi}_+={\varphi}_-$; in this case, the particle coincides with its own antiparticle. This is, for instance, what happens in the case of 
the photon, the particle responsible of  electromagnetic interactions. Because of this, the number of such particles may change after a reaction.

And what happens to matter (spin 1/2) particles? 
Let us consider Dirac equation, Eq.~\ref{eq3}. In 1937 \cite{ettore}, Majorana remarked that 
 in  
 principle, if the electric charge of the particle is zero, one could identify 
 \begin{equation}{\psi}_+={\psi}_-\end{equation} 
 using the above notation for the solutions, and Majorana's representation of gamma-matrices. Or to put it in simple words, these special spin 1/2 particles   would coincide with their own antiparticles - they would be matter and antimatter at the same time. Majorana  suggested that this might be the case for neutrinos or neutrons. 
 Neutrons were discarded because  of their magnetic moment \cite{racah}, 
  but for neutrinos this hypothesis is still considered plausible and corresponds to very topical questions.
%

\subsection{The structure of weak interactions} 
This observation could confuse a modern reader, accustomed to distinguish neutrinos from antineutrinos.
Obviously, this distinction is not based on the electrical charge of neutrinos, which is not there; so, it is worth going over how we arrived at certain beliefs, to clarify this conceptual point as much as possible.

In 1956, Lee and Yang \cite{ly1}
questioned parity conservation for weak interactions and shortly after the experiment of Wu \cite{wu}
proved the correctness of this supposition. In the next year (1957)  independently Landau, again
Lee and Yang, and Salam suggested that neutrinos could have a chiral type of interaction \cite{abd,dau,ly2}. 
Along with the  assumption that their mass is small, this implies that the projection of the spin $\vec{\Sigma}$
in the direction of the momentum $\vec{p}$, i.e., the helicity  
\begin{equation}\mathcal{H}= \frac{ \vec{\Sigma} \, \vec{p}} {p}\end{equation}
is negative for neutrinos and it is positive for antineutrinos. 
Stating it in plain words: 
{\em If neutrinos are supposed to be massless, helicity distinguishes them from antineutrinos.}
This was tested by Goldhaber in 1958 in lab \cite{gold}, using ultrarelativistic neutrinos. 

Finally Sudarshan and Marshak \cite{sud}, and independently Feynman and Gell-Mann \cite{fg}, generalized the point by 
suggesting that weak charged currents have a polar-minus-axial-vector (i.e., $V - A$)  structure, which is 
chiral. The modern reader glimpses one of the main pillars of the standard model behind these positions, but rather than jump too far, it is useful to understand at this point what role neutrino masses play.

\subsection{Majorana neutrinos and weak interactions} 
Now, let us examine what happens if the neutrino mass  
is not exactly zero. 
This point was clarified with a bit of difficulty in the 
scientific literature and  has stimulated much  interesting results 
\cite{case,whitten,mgs,cuspo,sp,bar,doi,lfl,sv,bk1,bk2}.
We will discuss this at length later, for now we would like to limit ourselves to describing the position we have arrived at on the basis of modern physics.
We present here a very transparent argument, based on \cite{simon}: 

Bringing the particles to rest (that is, when the momentum is zero
$p=0$) the helicity 
would no longer be defined, we would have only the spin $\vec{\Sigma}$. From the point of view of the spin, 
the two neutral particles (neutrino and antineutrino) would differ only for the state of rotation; 
however, an elementary particle should remain unchanged under rotations, 
which would lead us to think that they are the same particle.\footnote{As is well known, we can think of a particle with spin as a sphere without structure, rotating about a vertical axis.  The image reflected in a vertical mirror is indistinguishable from the particle with the axis of rotation reversed. The two cases are distinguishable when we have not only spin but also momentum (and especially in the case where momentum cannot be eliminated, as is the case for a massless particle).} This is consistent with the hypothesis of Majorana.

In principle, we could still invoke some special law to distinguish neutrinos from antineutrinos also in the rest-frame, which would 
double the number of particles. 
But Occam's razor would suggest this is not 
the first case to consider, as there would be no need of doubling the number of particles.
So we are lead to consider seriously the hypothesis of Majorana, and to expect that, in
the  rest frame, the 
$V-A$ interactions would produce the two spin states equally.
(The other position, where the number of neutrinos is doubled in the rest system, is the hypothesis that neutrinos are Dirac particles.)

Finally, let us recall the important and well-known observation 
that neutrino oscillations \cite{cuspo,uspo,cruspo} 
cannot distinguish Majorana masses from Dirac masses \cite{sp}; we refer to \cite{osc-d} for a further discussion of the hypothesis on the oscillations that interest us in practice, i.e., those occurring in the ultrarelativistic regime.

The above analysis showed that due to the chiral structure of the interactions, 
when neutrinos are relativistic the characteristic effects of Majorana neutrinos 
are suppressed. 
Further supporting (formal) arguments are presented in appendix~\ref{nan}.


%

\bigskip
In summary, we can say that 
\begin{itemize}
\item the possibility that the neutrino and antineutrino coincide in the rest-frame--i.e., that 
Majorana's hypothesis is correct--does not formally contradict what we know;
\item most of the empirical knowledge we have about neutrinos concerns only the ultra-relativistic limit instead, 
and the characteristic manifestations of Majorana hypothesis are suppressed in this limit.  
\end{itemize}
It becomes interesting to understand even better the meaning of this hypothesis - we will discuss it in the next section - and to put it to the test,  in the way described in Sect.~\ref{s:e5}.

\section{Status of baryon and lepton number conservation laws}\label{s:e4}
We now return to the conservation laws that are part of the definition of what matter is. In recent times, evidence has accumulated that all individual leptonic numbers
\begin{equation}
\mathbf{L}_{\rm e}\ , \ \mathbf{L}_\upmu\ , \ \mathbf{L}_\tau
\end{equation}
are violated, as was discovered through the experiments such as K2K, T2K, NO$\nu\!$A and OPERA (see Tab.~\ref{aperta}) which observed the 
appearance of a lepton of a different type (aka `flavor' aka `family'). Instead, as already recalled, there is no empirical evidence that 
their sum, the total leptonic number, 
\begin{equation}\mathbf{L}=\mathbf{L}_{\rm e} + \mathbf{L}_\upmu + \mathbf{L}_\tau\end{equation} 
be violated. Majorana's hypothesis for the neutrino evidently violates the leptonic number $\mathbf{L}$ 
by two units, but as we have discussed, these effects disappear 
with the neutrino mass. We know from observations that neutrino masses are small; the experimental basis for this conclusion is briefly reviewed in the next section just for completeness (as the story is rather well-known),  
and their implications will be better discussed in the next section. The rest, and the main part of this section, will be devoted to completing the description of the theoretical framework.

\begin{table}[t]
\centerline{
\begin{tabular}{|c||c|c|c|c|c|c|}\hline
 & $\Delta ( \mathbf{L}_{\rm e}-\mathbf{L}_\upmu)$ & $\Delta ( \mathbf{L}_\upmu-\mathbf{L}_\tau)$ & $\Delta ( \mathbf{L}_{\rm e}-\mathbf{L}_\tau)$ &
 $\Delta ( \mathbf{B}-\mathbf{L})$ & observations  \\ \hline 
 $\nu_\upmu\to \nu_{\rm e}$ & $+2$ & $-1$ & $-1$  & 0 & \cite{t2k,nova} \\ \hline
  $\nu_\upmu\to \nu_\tau$ & $-1$ & $+2$ & $-1$  & 0 & \cite{opera,sk,deepc} \\ \hline
\end{tabular}}
\caption{\small\sf  Selection rules of the observed violations of exact global symmetries of the standard model.}\label{aperta}
\end{table}

\subsection{Neutrino oscillations and the evidence of neutrino masses}\label{sec:cg}

The only experimental basis for stating that neutrinos have mass is the extensive evidence of neutrino oscillations. 
The existence of this phenomenon was hypothesized by Bruno Pontecorvo in 1957 \cite{cuspo,cuspoo}; 
at that time it was called ``virtual transition'', borrowing the terminology introduced for neutral $K$ mesons \cite{pais}, and 
it was (incorrectly) believed that neutrinos transformed into antineutrinos.\footnote{E.g., we read in \cite{cuspo} the words: 
{\sc if the conservation law of neutrino charge would not apply, then in principle neutrino $\to$  antineutrino transitions could take place in vacuo.}}
Its description was refined in the following years, leading, 10 years later, to the modern theory years later \cite{bruno}
that exploits the concept of leptonic mixing \cite{sak0,sak1,sak2}. 
A further important theoretical ingredient for the interpretation of the data is the ``matter effect''  \cite{w,ms}, attributable to 
an additional phase for $\nu_{\rm e}$ and $\bar{\nu}_{\rm e}$ due to weak interactions with electrons in ordinary matter. 

The first observational substantiation of oscillations was obtained by comparing the solar neutrino data \cite{home,kam,gal,sag} 
with the theory of  their production in the Sun \cite{bahc}. 
The results of \cite{snonc} allowed  to verify the interpretation in an almost model-independent way. 
Modern measurements, including those of \cite{sks} and \cite{borex}, together with KamLAND's results with reactor antineutrinos \cite{kl} 
allowed us to progress and 
measure precisely the relevant parameters, discovering that the electron neutrino is mainly the lightest component 
of a neutrino mass doublet - a result based on the attestation of ``matter effect''.

A second group of observational evidence is given by studies of atmospheric neutrinos. A decisive contribution is attributed to the Kamiokande experiment \cite{kama1,kama2} which evolved into the Super-Kamiokande experiment \cite{skk}. The results are consistent with many other observations of atmospheric neutrinos, beginning with those of MACRO \cite{macro}, of Soudan-II \cite{soud} and others. Again, verifications have been performed under controlled conditions, in particular those obtained thanks to the artificial beams produced in the accelerators \cite{k2k,t2k,nova,opera} (more on this just below).
In addition, further useful information 
has been obtained thanks to precision measurements in the \cite{daya,reno,chooz2} reactors. It is important to investigate the mass spectrum of neutrinos in more detail, to see whether it resembles the mass spectrum of charged fermions or not. 
At the moment we only have clues in favour of this option \cite{ind1,ind2,ind3,ind4};
the statistics that will be collected with the next generation of very large detectors 
 \cite{hk,juno,dune} will provide us conclusive answers. 
(We will return to the  meaning of the mass spectrum later.)

\subsection{Global symmetries in the standard model}
Let us consider ``standard model'' of elementary interactions,  based
on gauge symmetry 
SU(3)$_c\times$SU(2)$_{\mbox{\tiny L}}$ $\times$U(1)$_{\mbox{\tiny Y}}$ \cite{gws1,gws2,gws3}
and with the three families of quarks and leptons \cite{anom1,anom2,renom}. 
It is well known that 
it presents several  `accidental symmetries' that imply the conservation laws of  
$\mathbf{B}, \mathbf{L}_{\rm e},\mathbf{L}_\upmu,\mathbf{L}_\tau$, and of course also of their linear combinations such as {\bf L}, 
that are valid at the leading (perturbative) level. 
On the other hand, as just mentioned, the conservation of individual leptonic numbers is not 
compatible with some experimentally observed facts. 

It is even more interesting to observe that the symmetries associated to the specific conservation laws 
$\mathbf{B} -  \mathbf{L},$  $\mathbf{L}_\upmu- \mathbf{L}_\tau$ and $\mathbf{L}_\upmu- \mathbf{L}_{\rm e}$ 
are {\em exact} symmetries instead, i.e., free from quantum anomalies, while {\bf B} and {\bf L}, taken alone, are anomalous \cite{gan}. 
Thus, see again Tab.~\ref{aperta}, various symmetries that would be expected to be exact are violated, and 
the only (presumedly exact) symmetry in the standard model of which we do not know any exception at the moment is
\begin{equation}
\mathbf{B-L}
\end{equation}
It is useful to note that a Majorana mass term of neutrinos would mean its violation. 

Note in passing that the standard model does predict the existence of baryon number violation manifestations \cite{qnn} 
- through non-perturbative phenomena above the electroweak scale, called {\em sphalerons} - but they are not sufficient to justify the excess of 
baryons in the cosmos \cite{nonan}. This adds interest in physics beyond the standard model related to still unobserved global number violations.

\subsection{Standard model and Majorana neutrinos}

At first glance, it would not seem so easy to write a Majorana mass term for neutrinos in the standard model. For example, we can form a Majorana spinor    
$\bm\chi=\bm\nu_{\mbox{\tiny L}}+ \bm\nu_{\mbox{\tiny L}}^*$ with the ordinary neutrinos $\bm\nu_{\mbox{\tiny L}}$ and then include 
 in the Lagrangian density also the following Majorana mass term (see the appendix for details)
%
\begin{equation} \label{conci}
-\frac{m}{2} {\bm\chi^t}\, \mbox{\textgamma}^0 {\bm\chi} =-\frac{m}{2} \bm\nu_{\!\mbox{\tiny L}} ^t \, \mbox{\textgamma}^0\,  \bm\nu_{\mbox{\tiny L}} 
+\mbox{hermitian conjugate}
\end{equation}
but this term violates a gauge symmetry, the hypercharge $Y$, by 1 unit.\footnote{We use the normalization 
$Q=T_3+Y$, so that $Y=-1/2$ for the neutrino field.} 
On the other hand, this symmetry is broken spontaneously, and with this consideration in mind we are led to write   
the following term which is a perfect gauge invariant
\begin{equation}
 \bm\ell^t_{\mbox{\tiny L}} \, \mbox{\textepsilon}\, \bm{H}
 =
\left( \bm\nu_{\mbox{\tiny L}} \, ,\, \bm{e}_{\mbox{\tiny L}} \right) 
\left(
\begin{array}{cc}
0   &  1 \\
-1  & 0  
\end{array}
\right) 
\left( 
\begin{array}{c}
0  \\
v+\frac{\bm{h}}{\sqrt{2}}
\end{array}
\right) = v\, \bm{\nu}_{\mbox{\tiny L }}+\mbox{interactions}
\end{equation}
where the higgs doublet  $\bm{H}$ is given in the physical gauge,  
the expectation value is $v=174$~GeV and $\mbox{\textepsilon}=i\sigma_2$ is the invariant matrix of SU(2)$_{\mbox{\tiny L }}$.
This term behaves just like the spinorial field under Lorentz transformations, so we can use it to form a term 
of the Lagrangian density of the type
\begin{equation} \label{d5}
\delta\mathcal{L}=
-\frac{1}{2\, M} ( \bm\ell^t_{\mbox{\tiny L}a}\,  \mbox{\textepsilon}\, \bm{H})  \mbox{\textgamma}^0_{ab}  (\bm\ell^t_{\mbox{\tiny L}b}\, \mbox{\textepsilon}\, \bm{H})  +\mbox{hermitian conjugate}
\end{equation}
After spontaneous symmetry breaking, this term reproduces  that 
in Eq.~\ref{conci}, therefore yielding Majorana masses. Thus we identify
\begin{equation}m=\frac{v^2}{M}\approx 50\,\mbox{meV} \times \frac{6\! \times\! 10^{14}\,\mbox{GeV}}{M}\end{equation}
a relation showing that the neutrino 
mass values $m$, which have been discovered by means of the neutrino oscillation phenomenon, correspond to very large masses $M$.
We note that this mass scale  strongly differs from $v = 174$ GeV, the electroweak mass scale, and is smaller than the Planck mass: a valuable indication of new physics.

\subsection{Theoretical remarks}

As we see, it is allowed to consider Majorana neutrinos in a manner where  the gauge symmetries of the standard model are respected, provided that operators of dimension 5 (i.e., operators that are not renormalizable) are included. Proceeding in this way, the possible question of the origin of these masses is  postponed to a subsequent, renormalizable formulation of an extension of the standard model, compatible with it.

There is no shortage of attractive theoretical options.
The operator in Eq.~\ref{d5} was first considered by Minkowski \cite{min} in a model that includes heavy right-handed neutrinos $\nu_{\mbox{\tiny R}}$, whose quantum fluctuations are suppressed by the inverse of the mass of such particles. Later, it was noted that such a situation is common to several models \cite{yana,glash} including those with extended gauge symmetries \cite{georgi,moha}. In some of these  grand unified theories (GUT), {\bf B$-$L} becomes a (spontaneously broken) gauge symmetry \cite{pat1,pat2,moha}, while in others the presence of new particles (including the heavy neutrinos $\nu_{\mbox{\tiny R}}$) offers new possibilities to explain the origin of the baryon asymmetry in the cosmos  \cite{fuku}. 

For the discussion of observable effects in the lab (that we will conclude in the next section) 
the most convenient language is that of effective operators, as argued in \cite{wei,zee}. The operator shown in Eq.~\ref{d5}, describing Majorana masses and that violates {\bf L} and also {\bf B$-$L}, is unique and is suppressed only by a single power in the mass scale of the new physics. 
In other words, this is enough to endow the standard model with small Majorana masses of the ordinary neutrinos. 
Proton decay arises with operators of dimension 6, pure baryon number violation phenomena arise with operators of dimension 9, etc. 
In this scheme, small violations of global numbers are attributed to new physics, which is fully manifested at scales different 
from those of the standard model. Note that strictly speaking none of these effects have been experimentally verified; however, we have indications that neutrinos have mass, and we know that their values are small, which suggests the presence of the dimension 5 operator just discussed.

Summarizing 
we  reached the following conclusions:
\begin{itemize}{\sf
\item[$\star$] 
the structure of the standard model does not contradict and in fact revives Majorana's hypothesis:
 neutrinos,  sole among all matter particles, are likely to be  their own antiparticles; 
\item[$\star$]  neutrino masses are expected to be very small and their value can be regarded as a special observational window on the physics far beyond the standard model itself.\footnote{However, it should be noted that comparison with models remains unavoidable: e.g.,  if the $\nu_{\mbox{\tiny L}}-\nu_{\mbox{\tiny R}}$ couplings are small,  it is possible to obtain small Majorana masses with lighter $\nu_{\mbox{\tiny R}}$.}}
\end{itemize}
These conclusions are stated in the context of the extended 
standard model. 

%

\section{Matter creation in lab?}\label{s:e5}
In the light of the above considerations, it is more interesting than ever to test the hypothesis that neutrinos have a small Majorana mass, which would make them without equal among matter particles. In fact, neutrinos would be the only elementary particles known to be both matter and antimatter, a hypothesis not only compatible with current theoretical thinking but even plausible in this context, as seen just above. 

Over the years, it has become increasingly clear that the most promising way to validate this hypothesis is to search for a rare nuclear transition in which two neutrons turn into two protons and two electrons, symbolically,
\begin{equation} \label{crea}
\rm 2 n \to 2 p + 2 e^-
\end{equation}
which, in practice, can be done by studying some nuclear transitions such as 
${}^{76}\mathrm{Ge}\to {}^{76}\mathrm{Se}+ \rm 2 e^-$,
${}^{130}\mathrm{Te}\to {}^{130}\mathrm{Xe}+ \rm 2 e^-$,
${}^{136}\mathrm{Xe}\to {}^{136}\mathrm{Ba}+ \rm 2 e^-$  and others, 
see \cite{touschek} \cite{tete,gege,xexew,xexe} for early works and
\cite{ccc}, \cite{alal,bbr,bbr3,bbr4} for a few recent reviews.\footnote{There is also a  
useful review of the numerous false starts \cite{vol}, which illustrate the great desire to be able to measure the aforementioned transition and which above all suggest that one should be cautious before accepting a discovery.} 
The most important fact for our discussion is  that this transition must be thought of, in light of current theories,  
as 
\begin{itemize}{\sf
\item[$\star$] a violation of the symmetry $\mathbf{B-L}$ due to the 
net creation of two matter particles (electrons)}
\end{itemize}
This is the reason why in the rest of this discussion we can refer to this process as `matter creation'. Of course, this is exactly the same process that we use to call ``neutrino-less double beta decay'' or in similar manners. However, in this discussion I would like to emphasise its relevance to current (and future) 
understanding of what matter is according to particle physics, and a new terminology suited for this sake is necessary.

Before outlining the  actual connection between Majorana neutrino mass and the process of Eq.~\ref{crea}, 
let us briefly recall how the discussion of this important process has evolved over time, examining the most important conceptual changes.

\subsection{Historical introduction}
Immediately after Majorana, Racah discussed the interactions of a Majorana particle, showing that weak vector interactions would have physical manifestations distinguishable from those of a Dirac particle \cite{racah}. Such a possibility was investigated in 1954, when Davis enquired whether 
the particle $\bar\nu$ produced in the reactors would trigger  ${}^{37}{\rm Cl}+\bar\nu \to  {}^{37}\rm{Ar}+e^-$: 
the answer turned out to be negative \cite{dacu}. 
Later in the literature, the term ``Racah chain'' was used to refer to a sequence of processes such as this; see~\cite{uspo}, which reconstructs the story with great accuracy. 


Instead, the specific process of matter creation of Eq.~\ref{crea} was first discussed by Furry in 1939 \cite{furry}, who was inspired by the process of 
``double beta disintegration'' previously discussed by  Goeppert-Mayer\footnote{We quote the beginning of the abstract: 
{\sc the probability of simultaneous emission of two
electrons (and two neutrinos) has been calculated.} Thus, this is a weak process in which the number of leptons remains unchanged,  that using modern notations, can be indicated schematically as ${\rm 2 n \to 2 p + 2 e^-}+2\bar{\nu}_{\rm e}$, compare with Eq.~\ref{crea}; today is often denoted  `double beta with two-neutrinos' and it has been measured for various nuclei.}  \cite{maria} and of course by Majorana's ideas. 
However, it is important to repeat that the theory of weak interactions  differed in crucial manner from the one we have today. 
If Fermi's vector interaction had been correct in the strict sense, as originally thought, the Majorana hypothesis would have had a much greater effect on the process of creation of matter. (This would also had been true for any other interaction except for a chiral theory). 
This was Furry's expectation.\footnote{The technical discussion on  App.s~\ref{nan}-\ref{sth} applies to the current theory of weak interactions instead.}

\bigskip

Table~\ref{bult}  illustrates the fortunes of the works  \cite{ettore}, \cite{maria}, \cite{racah}, \cite{furry} and \cite{case}
over time; the last paper is included for reasons 
explained in  App.~\ref{tooe}.
Even if it is indisputable that the number of scientific publications is growing faster and faster, it seems difficult to deny a fact, that the greatest interest in Majorana's work arose in relatively recent times. 
In order to understand better the meaning of this late outburst of interest, 
let us begin by identifying three main periods: 
\begin{itemize}
\item the period of theoretical novelty, 
in which the hope of observing very large effects was highlighted by Furry 
and Majorana's neutrinos were actively discussed by scientists such as Stueckelberg, Kemmer, Touschek, Pauli, Harish-Chandra, Michel, Yang, Tiomno, {\em etc.};
\item the period begun by  the understanding of the ($V-A$) \underline{structure of weak interactions}, where, on the other hand, there was an exaggeration in the opposite direction, and the doubt was cast whether Majorana's theory should be dismissed - more on App.~\ref{tooe};
\item the current period, where the stress shifted from the concept of 
Majorana's neutrinos to the more precise one of \underline{Majorana neutrino masses}, subsequently
including their meaning in the context of the extensions of the standard model.
\end{itemize}
(Compare with the much more detailed tables in Pontecorvo's review work \cite{uspo}, already cited, and with the  valuable 
historical study \cite{silvia}.
The introduction to the book \cite{klapi} also has a table comparable with Tab.~\ref{bult}; the main difference 
 is that it lacks a discussion of the second, crucial period, which we present in this section and complete 
 in App.~\ref{tooe}.\footnote{On the other hand, the book includes a collection, albeit incomplete, of many relevant articles useful for the discussion we are interested in.})

 \begin{table}[t]
\centerline{
\setlength{\tabcolsep}{0.26em}
\begin{tabular}{c|c|c|c|c|c|c|c|c|c}
\hline\footnotesize
                                         &  $'30$s  & $'40$s & $'50$s & $'60$s & $'70$s & $'80$s & $'90$s & 2000 & 2010  \\ \hline\hline
\footnotesize Majorana \cite{ettore} & 3 & 3 & 8 &  
5 & 17 & 
43 & 67 & 159 & 745   \\  \hline  
\footnotesize Goeppert-M. \cite{maria} & 2 & 2 & 6 &  
0 & 0 & 
18 & 19 & 41 & 221   \\  \hline  
\footnotesize Racah  \cite{racah} & 2 & 1 & 6 &  
1 & 6 & 
19 & 16 & 31 & 133   \\  \hline 
\footnotesize Furry  \cite{furry} & 0 & 2 & 6 &  
0 & 1 & 
25 & 30 & 71 & 351   \\  \hline 
\footnotesize Case  \cite{case} & - & - & 1 &  
10 & 10 & 
36 & 43 & 34 & 36   \\  \hline 
 \hline
\footnotesize theory & \small $\nu$ &\small \bf B\it,\,\bf L & \footnotesize \em V-A,\,\em SU(2) & \footnotesize $\!$SM,\,oscill.$\!$ & \footnotesize SM  &  \footnotesize GUT,\,$\nu_{\mbox{\tiny $\odot$}}$ & \footnotesize susy &  \footnotesize glob.anal.  & \footnotesize cosm.   \\ [-0.2ex]\hline
\footnotesize exp.\ \& obs. &  \footnotesize n,e$^{\mbox{\tiny +}}\!,\upmu$ &  \footnotesize $\pi,K$  & \footnotesize $\nu$,\,\em V-A & \footnotesize $\nu_{\mbox{\tiny $\odot$}}$\,anom. & \footnotesize SM,\,$\nu_{\mbox{\tiny $\odot$}}$  & \footnotesize $W,Z^0\!$,$\nu_{\mbox{\tiny atm}}$ &  \footnotesize oscill. & 
\footnotesize oscill.  & \footnotesize higgs,$\,$cosm.  \\ \hline
\end{tabular} }
\caption{\small\sf  Number of papers per decade citing few seminal articles,  including Majorana's, 
from their appearance to the present day. In the last two lines, a (incomplete) list of theoretical and observational facts that are of major relevance  for current discussion. From \url{https://inspirehep.net/} [Dec.~2020]. 
\label{bult}}
\end{table}

Let us add a few comments to put this historical examination in the perspective of the discussion we are interested in.
Rather plausible theoretical ideas  lead us to think that neutrinos have small masses of the type considered by Majorana. 
Furthermore, the existence of neutrino oscillation phenomena, predicted in  \cite{bruno,ms,w},  is nowadays widely 
recognised, also by 2015 Nobel prize in physics and it has been 
verified in controlled experiments, in particular those mentioned in Tab.~\ref{aperta}. 
This leads to a firm conclusion that neutrino masses are not zero.\footnote{From this point of view, it should be stressed that an activity whose significance in theoretical physics might seem apparently modest, that of the global analysis of neutrino oscillation data, acquires instead an enormous importance since '90s, and this is the reason why we mention it explicitly in the table above: this is the only way to date to measure the neutrino mass parameters. 
See e.g., \cite{elio1,elio2} that witness a continued effort over a period  of about 25 years.} 
Thus, a combination of theoretical and experimental considerations let us to think, today, 
that the rate of the decay process first considered by Furry (that we regard as a matter creation process) 
is controlled by the value of Majorana's neutrino mass.

These considerations  justify the enduring efforts in testing this process in laboratory.
In fact, several important experiments of this kind have already been carried out and new efforts are underway, see \cite{kz,cu,cp,ge} for a few recent experimental searches and \cite{simon,bbr,bbr3,bbr4} for reviews.

%
%
%

\subsection{What we know on the relevant Majorana neutrino mass}
The structure of the standard model is compatible with the idea of very small Majorana neutrino masses, to be attributed to new physics, at much larger energy scales. This suggests that there are other physical manifestations, but that they are not (easily) detectable at low energies, and conversely, makes it even more important to carefully plan those  measurements that can be actually performed, based on the hypothesis that {\em the leading reason  of the electron/double beta creation rate of Eq.~\ref{crea} is Majorana neutrino masses.} 
A correct evaluation of nuclear matrix elements is essential to obtain a prediction of the rate, and as it is well-known this implies  uncertainties, not yet fully clarified. But the main reason of uncertainty in the prediction remains simply that due to the value of the Majorana mass that is relevant for this process. Therefore, we will conclude by addressing the quantitative aspects, which are brought into play by these theoretical assumptions.

Suppose that the three ordinary neutrinos  have 
Majorana mass. We will have the following Lagrangian density\footnote{The complex numbers $m_{\ell \ell'}$ form a symmetric matrix; the proof is in Eq.~\ref{quibus}. Its decomposition 
can be presented as a relation among matrices: 
 $\hat{m}  =\hat{U}^* \  { \mbox{diag}\small(\tiny m_1,m_2,m_3) }
\ \hat{U}^\dagger $.}  
\begin{equation}
\delta\mathcal{L}=-\frac{m_{\ell \ell'}}{2} \bm\nu_{\mbox{\tiny L}\ell}^t\,  \mbox{\textgamma}^0 \,  \bm\nu_{\mbox{\tiny L}\ell'} + \mbox{h.c., where }
m_{\ell \ell'} = U_{\ell i}^*\  m_i\ U_{\ell' i}^* 
 \end{equation}
 where the indices are $\ell,\ell'={\rm e},\upmu,\tau$ and point to its components in `flavor space';
 $m_i\ge 0$  are the neutrino masses and $U_{\ell i}$ is the leptonic mixing matrix. 
 As it is evident from the selection rule on 
 the electronic lepton number $\bf L_{\rm e}$, the quantity that matters for the transition of Eq.~\ref{crea} is only the matrix element 
 $\mee\equiv m_{\upbeta\!\upbeta}$, that can be chosen to be real and positive; see again App.~\ref{sth}.
  
  The neutrino oscillation experiments (see Sect.~\ref{sec:cg})
  have opened for measuring very precisely the absolute values of $U_{\ell i}$ and the absolute value of the differences $m_i^2-m_j^2$. 
The sign of the latter quantity is known precisely for the two masses closest to each other (called $m_1$ and $m_2$) while we do not know for sure whether $m_3$ is larger or smaller than the other two masses. To be well defined, we will focus on the case $m_1<m_2< m_3$. The experimental evidence, albeit weak, is currently in favour of this position. In this way, the mass spectrum of neutrinos more closely resembles that of charged fermions, and is therefore called the `normal mass spectrum' (aka `mass hierarchy' aka `mass ordering').

\begin{figure}[t]
\includegraphics[width=0.47\textwidth]{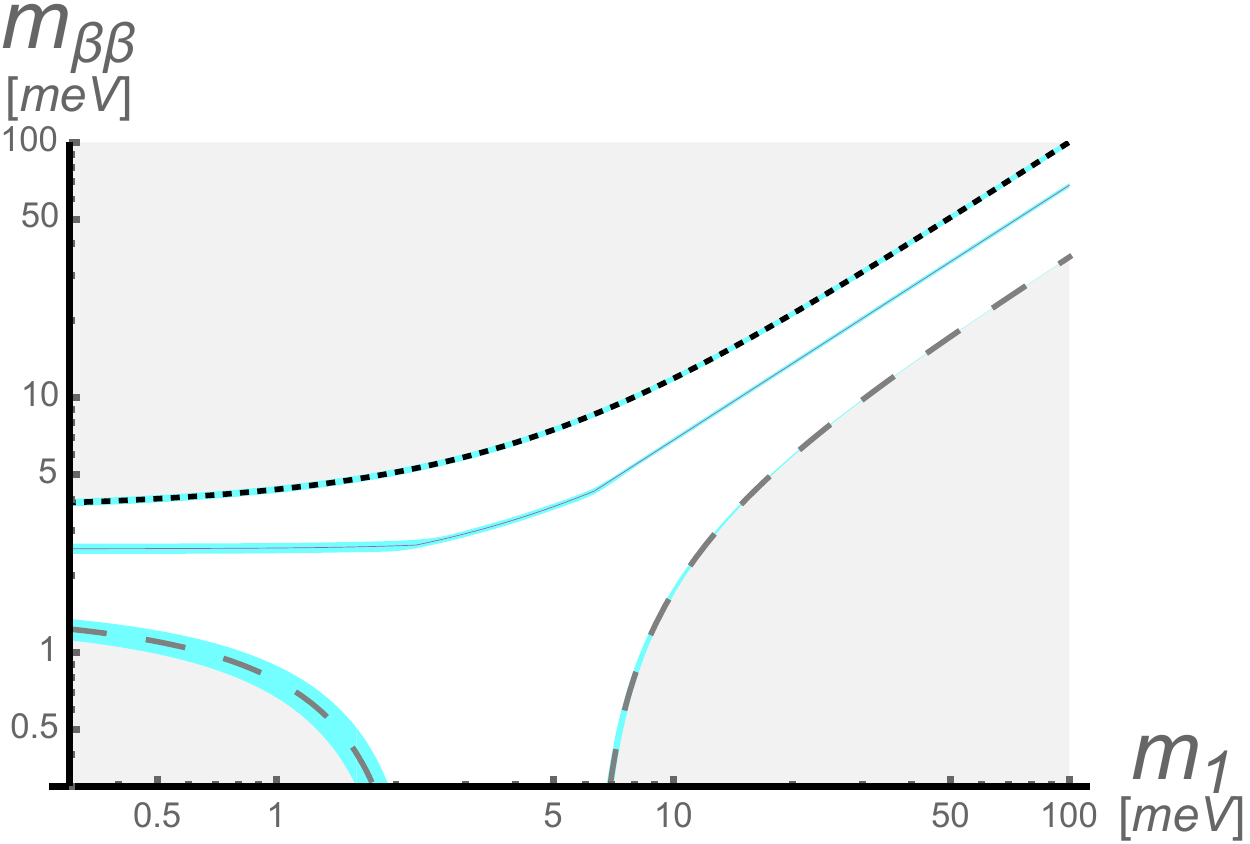}\hfill\includegraphics[width=0.47\textwidth]{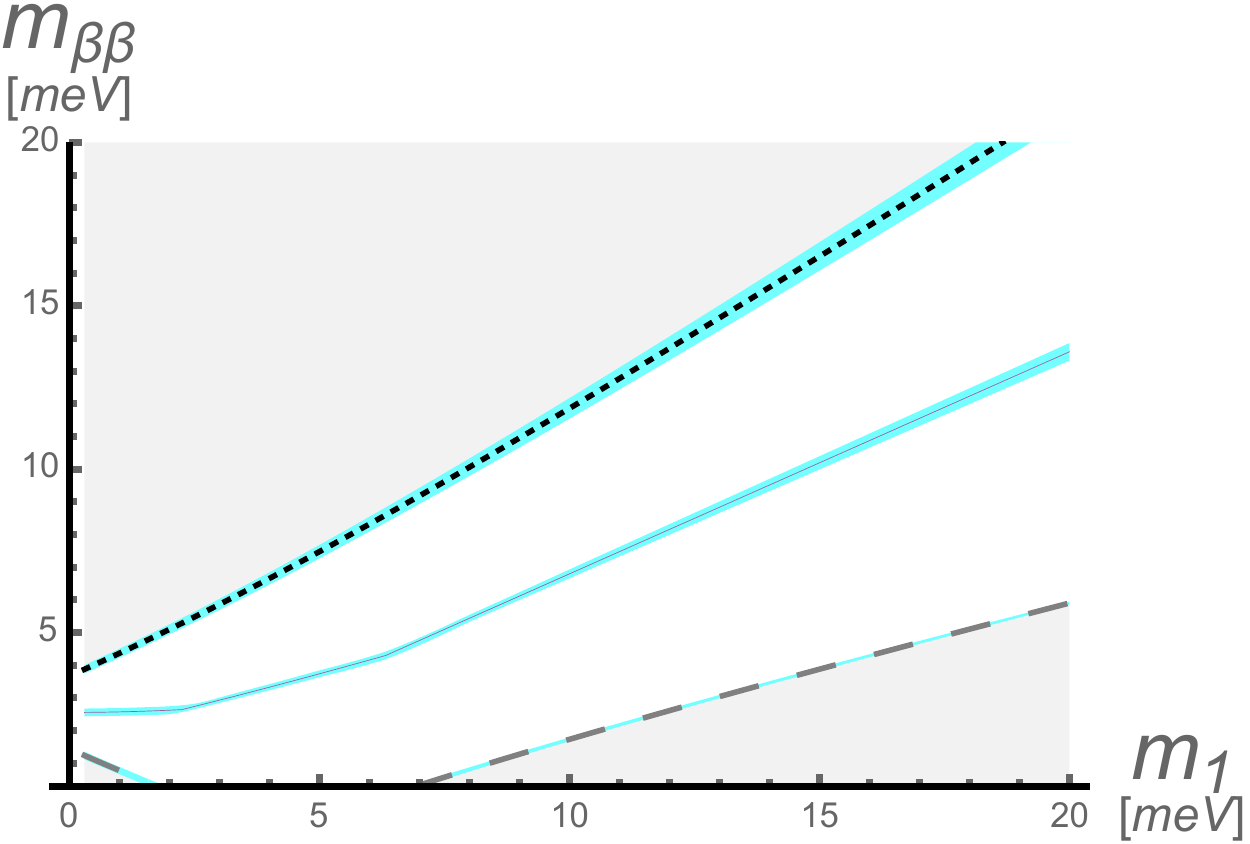}
\caption{\sf\small 
Allowed region for the parameter $m_{\upbeta\!\upbeta}=\mee$ (white) assuming normal mass spectrum. On the left, log-log graph, on right,
zoom in the region of the smallest $m_1$ somewhat favored by 
cosmology (see text and Fig.~4 of \cite{sim1}).
 The central curve is the average value.
 The colored bands indicate the uncertainty regions (1 sigma).
Gray regions are incompatible with observed 
neutrino oscillations.}
\label{lava}
\end{figure}

The real trouble is that 
\begin{quote}
{\em the oscillations   probe neither the phases of $U_{\ell i}$ 
  nor the  mass of the lightest neutrino $m_1$.}
  \end{quote}
 This implies that the parameter $\mee$ can range from 0 to the maximum value allowed by experimental measurements. 
 The theory of fermion masses (and in particular the one of neutrino masses) is not sufficiently developed to be reliable, and 
  it is difficult to obtain further useful information. 
  
To proceed on an empirical basis, the following steps were taken.
In order to address the  problem  of the phases (sometimes called ``Majorana phases'')
 it was proposed in \cite{vis99} 
 to consider the maximum and minimum value of $\mee$, 
by varying the unknown complex phases, obtaining the formulas 
\begin{equation}
\begin{array}{l}
\\[-4ex]
\displaystyle \mee^{\mbox{\tiny max}}  =\sum_{i=1}^3 | U_{\rm e\it i}^2| \  m_i  \\[3ex]
\displaystyle \mee^{\mbox{\tiny min}}  = \mbox{max}\!\left\{\,   2 \, | U_{\rm e\it i}^2| \,  m_i -\mee^{\mbox{\tiny max}}\,  ,\,    0\,  \right\}
\end{array}
\end{equation}
see Fig.~\ref{lava} for the standard presentation.\footnote{The average value
{\rm $(\mee^{\mbox{\tiny max}} +\mee^{\mbox{\tiny min}})/2$} showed in Fig.~\ref{lava} has a simple analytical expression;
e.g., for large $m_1$ it is  just {\rm $|U_{\rm e 1}^2| \  m_1 $}, in which the mixing parameter  
$U_{\rm e 1}$  is almost perfectly known and $m_1$ is not at all.}

The second problem, namely the fact that the mass of the lightest neutrino is unknown, can  be tackled in two ways 
\begin{itemize}
\item In  principle, by means of extremely precise measurements of the absolute neutrino mass $m_{\nu_{\rm e}}^2= \sum_i |U_{{\rm e} i}^2| m_i^2$ (a parameter discussed in \cite{sh1,sh2,sh3,vis00,alew})  in the laboratory. Current experiments indicate $m_{\nu_{\rm e}}<0.8-1.1$ eV at 90\% and the sensitivity will reach 200 meV in the near future
\cite{katrin}. The current limit directly translate into $m_1<0.8-1.1$ eV.
\item From cosmology measurements that probe  $\Sigma=m_1+m_2+m_3$ as suggested in~\cite{lisic}. Even if they are based on a cosmological model that is conceptually complex and still under verification,  
the sensitivity of the most recent measurements is just impressive: see e.g.~\cite{cosmmeas1,cosmmeas2,cosmmeas3}. 
Proceeding as in \cite{sim2,stef} we get $\Sigma<100$ [150] meV at 1 [2]$\sigma$ that translates into 
$m_1<20$ [40]~meV.
\end{itemize}
An empirical approach based on cosmological measurement 
results in a upper bound of $\mee$ of a few 10 meV. 
This result, obtained and discussed  in \cite{sim2,stef,sim1,lisiadd}, reinforces the view stated in \cite{simon}, that it will be a real challenge to be able to observe the electron creation process of  Eq.~\ref{crea}. In particular, we can already conclude on this basis that detectors with a mass of many tonnes, and able to operate under stable conditions for many years, will be needed in order to have a chance of detecting the transition due to the Majorana mass of ordinary neutrinos, as argued in \cite{simon}. 
%

%




\section{Summary and discussion}

In this paper, we have retraced some important steps in the understanding of what matter is, according to modern particle physics. We have shown how Majorana's ideas are rooted right at the heart of the discussion about the question whether it is possible to create particles of matter. Trying to summarise, we could say that the search for the value of Majorana's neutrino mass is gaining momentum.\footnote{A side comment concerns some basic techniques proposed in \cite{ettore} and illustrated in the appendix, which still retain a certain practical usefulness.}

We reiterate that we have taken care to touch the main historical lines, but it is important to be aware that at any given time there are  forward-looking ideas, rearguard discussions, and even missteps. Obviously, it is not easy to correctly identify one's position in the present, i.e.~at the moment when history is being made, but we believe that having a correct perception of the road already travelled can help to keep the bar in the right direction.  In this spirit, and by way of summary, we provide in Tab.~\ref{ciu} a schematic description of the evolution of ideas about what matter is.

We stress that this review has no claim to completeness, and that we have left aside many interesting and widely discussed topics in the scientific literature. Just as a specific example, let us consider the idea that the transition  proposed by Furry, Eq.~\ref{crea},  is {\em not due} to the Majorana masses of neutrinos, but rather to some other cause. This hypothesis has also been examined and thoroughly discussed: for the first time by Touschek in 1948 \cite{touschek} (see also \cite{serpe} and \cite{silvia}), then just after the discovery of the chiral structure of weak interactions \cite{fei}, again after the understanding of neutrino oscillations \cite{pont}, and many other times recently.  It is not possible to rebut it on an empirical basis, and current theories are not in a position to exclude it absolutely. Nevertheless, it does not seem possible to argue that it has the same interest as Majorana's hypothesis, especially in the context of motivated extensions of the standard model, and unless the scale of new physics turns out to be unexpectedly low.

To conclude this discussion, we would like to express an impression somewhat conveyed by this brief historical review: while it is certainly important (essential) to have new experimental facts to discuss, it seems at least as important to have clear and well-defined theoretical ideas to enable real progress,
and it is essential to be able to acknowledge and value them. Maybe not a very strong point, but useful to work on. 


Finally and  quite importantly, it must be said that there is no shortage of tasks for the future, 
and most of them are quite demanding. In fact, efforts will be needed, devoted
1)~to consolidate inferences about the mass of the neutrino (by means of cosmology and laboratory investigations);
2)~to continue and to refine the experimental search for the nuclear transition in which two particles of matter (=electrons) are created;
3)~to make progress in the evaluation of the nuclear matrix elements related to this transition, quantitatively estimating the errors in expectations; 
4)~and last but not least, to develop convincing and principled extensions of the standard model,  that allow us to obtain useful and reliable indications on the expected value of the Majorana neutrino mass.

\begin{table}[t]
\centerline{\small
\setlength{\tabcolsep}{0.29em}
\begin{tabular}{r|r |||||||l|l}
\hline
matter & salient & valid & reason for \\
 components &  features  & until & inadequacy\\ \hline
atoms & species, mass & [1838] 1909 & [atoms of electricity] electron \\
 nuclei \& e$^-$ & charge, mass, spin & [1930] 1956 & [Fermi th.] neutrons \& neutrinos\\
p, n, e, $\nu_{\rm e},$ $\upmu$, ... & {\bf B}, {\bf L}$_{\rm e}$, ...,  " , " & [1961] 1968 & [standard model] quarks\\
quarks \& leptons &  {\bf B-L}, {\bf L}$_{\rm e}$-$\mathbf{L}_\upmu$, {\bf L}$_{\rm \upmu}$-$\mathbf{L}_\tau$, " , "  
& [1962] 2010 & [lepton mixing] appearance exp. \\ 
quark/antilepton & {\bf B-L}, " , "  & [1937]\ \ \ \ \  ? & [Majorana mass] 2n$\to$2p+2e \\
fermions &   mass, spin & [1977?] ??? & [supersymmetry?] ??? \\ \hline
 \end{tabular}}
 \caption{\sf\small Table of elementary components of matter according to different models, successive in time. Left columns: 
 brief description. Right ones: shortcomings of the given model. 
  Dates and information [in brackets] refer to mainly theoretical considerations. 
 The current model of what matter is 
 is the one in the penultimate line, since the Majorana mass hypothesis is very reasonable, but does not yet enjoy observational evidence; to date, the limitations of the last model are only speculative.}\label{ciu}
 \end{table}

%
%


\newpage
\appendix

\section{On Majorana spinors}

In this appendix, we cover several technical aspects concerning Majorana spinors \cite{ettore}.
Here (as in the rest of this paper) 
we use consistently Majorana's representation of the gamma-matrices just as advocated e.g.,  
in \cite{lfl}.
 For further   discussions see~\cite{simon,osc-d}. 
After introductory matter, Sect.~\ref{maj} and~\ref{nan}, we present  in Sect.~\ref{sth}
the definition of the key parameter, 
 $\mee$. A major obstruction for a useful discussion of Majorana's neutrino mass, 
 which  persisted for more than twenty years, 
is  examined critically in Sect.~\ref{tooe}.
  
\subsection{Illustration of the usefulness of Majorana representation}\label{maj}
Thanks to the characteristic feature of Majorana representation of gamma-matrices, all
relevant $4\times 4$  matrices have simple properties under conjugation and 
transposition, given in Tab.~\ref{tbl}. This is rather useful in various manipulations.

\paragraph*{First example:} Let us begin showing the consistency of 
Majorana's condition for a spinor, 
\begin{equation}\label{mcd} \bm\chi=\bm\chi^*\end{equation} 
The generic Lorentz transformation  is parameterized
by three angles $\theta_i$ and three rapidities\footnote{This parameter is connected to the velocity by 
$\beta=\tanh\eta$,  
so that $t'=\mbox{\textgamma} (t+\beta x)=\cosh \eta\, t+ \sinh \eta\, x$.} 
$\eta_i$:
\begin{equation}
\begin{array}{c}
\delta \bm\chi=  \frac{i}{2} \left( \vec{\Sigma} \, \vec{\theta}  +  \vec{K} \, \vec{\eta}  \right) \bm\chi \\
\delta \bm\chi^*=  \frac{i}{2}   \left( -\vec{\Sigma}^* \, \vec{\theta}  -  \vec{K}^* \, \vec{\eta}  \right)  \bm\chi^*
\end{array}
\end{equation}
Since the spin and the boost matrices are both imaginary, as shown in Tab~\ref{tbl}, it is evident that $\bm\chi$ and $\bm\chi^*$
transform in the same way.

\begin{table}[t]
\centerline{\begin{tabular}{c|c|c|c|c|c|c|c}
& $\mbox{\textgamma}^0$ & $\vec{\mbox{\textgamma}}$  & $\mbox{\textgamma}^5$ & $\beta$  & $\vec{\alpha}$ & $\vec{\Sigma}$ & $\vec{K}$ \\ \hline 
hermiticity & + & $-$ & + & + & + & + & $-$ \\
reality & $-$ & $-$ & $-$ & $-$ & + & $-$ & $-$ \\
symmetry & $-$ & + & $-$ &  $-$ & + & $-$ & + \\
\end{tabular}}
\caption{\small\sf \sf Properties of gamma-matrices in Majorana representation.
We define as usual $\mbox{\textgamma}^0=\beta$ and $\vec{\mbox{\textgamma}}=\beta\, \vec{\alpha}$ from Dirac's Hamiltonian; 
$\mbox{\textgamma}^5=i\mbox{\textgamma}^0\mbox{\textgamma}^1 \mbox{\textgamma}^2\mbox{\textgamma}^3$ as chirality matrix;  
$\Sigma_{jk}=i \mbox{\textgamma}_j \mbox{\textgamma}_k$.
Generators of Lorentz's group: 
$\Sigma_1=\Sigma_{23}$, $\Sigma_2=\Sigma_{31}$, 
$\Sigma_3=\Sigma_{12}$ (spin) and 
$\vec{K}=i \mbox{\textgamma}^0 \vec{\mbox{\textgamma}}=i \vec{\alpha}$ (boost). The indices $i,j,k$ take the values $1,2,3$.}\label{tbl}
\end{table}

\paragraph*{Second example:}  As a second  example of application, let us consider the familiar 
Lagrangian density of a Dirac field $\bm\psi$ (recall that $\hbar=c=1$)
\begin{equation}
\mathcal{L}=\bar{\bm\psi}( i \mbox{\textgamma}^\upmu \partial_\upmu-m) \bm\psi =   {\bm\psi}^\dagger ( 
 i \partial_0 - i \vec{\alpha} \, \vec{\nabla}  - m \beta) \bm\psi  
\end{equation}
where $\upmu=0,1,2,3$, $\partial_\upmu=\partial/\partial x^\upmu$ 
and $(\vec{\nabla})^i= \partial^i$ with $i=1,2,3$. 
We would like to rewrite $\mathcal{L}$ 
with Majorana's fields, still using  Majorana's  representation of gamma-matrices. 
Thus, let us  express the Dirac  field as follows,  
\begin{equation}\bm\psi_a=\frac{ \bm\chi_a +i\, \bm\lambda_a}{\sqrt{2}} \mbox{ where } \bm\chi^*=\bm\chi,\ \bm\lambda^*=\bm\lambda\end{equation} 
which resembles closely the separation of a complex number into its real and imaginary parts. 
Various mixed terms appear, but we 
 note that, owing to anticommutativity of the fermionic fields, they 
are either zero or `surface terms' that can be omitted in the action $S=\int d^4x\, \mathcal{L}$,
\begin{equation}
\begin{array}{c}
\bm{\chi}^t \partial_0 \bm\lambda - \bm{\lambda}^t \partial_0 \bm\chi =\partial_0( \bm{\chi}^t \, \bm\lambda ) \\
\bm{\chi}^t \alpha^i \partial_i \bm\lambda - \bm{\lambda}^t \alpha^i \partial_i \bm\chi = \partial_i( \bm{\chi}^t \alpha^i \bm\lambda ) \\
\bm{\chi}^t \beta \bm\lambda - \bm{\lambda}^t \beta \bm\chi =\bm{\chi}^t \beta \bm\lambda - \bm{\chi}^t \beta \bm\lambda =0 \\
\end{array}
\end{equation}
Therefore, the result is equivalent to two decoupled lagrangian densities,
\begin{equation}
\mathcal{L}=\frac{1}{2} \bar{\bm\chi}( i \mbox{\textgamma}^\upmu \partial_\upmu-m) \bm\chi + 
\frac{1}{2} \bar{\bm\lambda}( i \mbox{\textgamma}^\upmu \partial_\upmu-m) \bm\lambda 
\end{equation}

\paragraph*{Third example:} 
From the last formula,  we read the Majorana mass term 
\begin{equation}
-\frac{m}{2}  \bar{\bm\chi}\, \bm\chi=-\frac{m}{2}  \bm\chi^t\, \mbox{\textgamma}^0\, \bm\chi
\end{equation}
a simple  expression that applies for Majorana's form of gamma matrices. 
Now we verify explicitly  that this term is non-zero, using the components. Let us consider 
a term in a Lagrangian density with  two spinors
\begin{equation}\label{quibus}
\bm\chi_a\, \mbox{\textgamma}^0_{ab}\, \bm\lambda_b = - \bm\lambda_b\, \mbox{\textgamma}^0_{ab}\, \bm\chi_a =  \bm\lambda_b\, (- \mbox{\textgamma}^0)^t_{ba}\, \bm\chi_a =
\bm\lambda_b\,  \mbox{\textgamma}^0_{ba}\, \bm\chi_a 
\end{equation}
where in the first step we have used anticommutativity 
of the spinors, which are quantized fermionic fields, and in the last passage we use  
antisymmetry of $\mbox{\textgamma}^0$, pointed out in the Tab.~\ref{tbl}. 
 This calculation might seem pedantic, however is not entirely useless. 
 E.g., consider one statement of \cite{mcl}: 
 {\footnotesize\sc``the scalar $\bar{\psi}\psi$ of the Majorana theory vanishes identically''}; 
when we replace $\bm\chi=\bm\lambda=\psi$ in Eq.~\ref{quibus}, we see that this statement has no valid basis and we can rebut it.\newline
  Finally, treating the fields 
  as matrices and applying the usual rules for hermitian conjugation, namely $(\bm{a} \bm{b})^*=\bm{b}^* \bm{a}^*$,
we can show that the same term is hermitian (real), 
\begin{equation}\label{quibus2}
(\bm\chi_a\, \mbox{\textgamma}^0_{ab}\, \bm\lambda_b)^* =  \bm\lambda_b^*\, (\mbox{\textgamma}^0_{ab})^*\, \bm\chi_a^*   
=  \bm\lambda_b\, (\mbox{\textgamma}^0_{ab})^*\, \bm\chi_a
=  - \bm\lambda_b\, \mbox{\textgamma}^0_{ab}\, \bm\chi_a
\end{equation}
In the second passage  we used the reality condition of Majorana spinors, in the third the fact that $\mbox{\textgamma}^0$ is imaginary. 
Then, applying the same manipulations as in Eq.~\ref{quibus}, we conclude the proof of hermiticity -- which implies the unitarity of time evolutor.

\paragraph*{Fourth example:} 

Let us consider the coupling of a  fermion of Majorana with an external 
electromagnetic field. In the case of an electrostatic interaction,  the hamiltonian density that describes the interaction  is 
\begin{equation}
\mathcal{H}=q\, A^\upmu_{\mbox{\tiny ext}} \bar{\bm{\psi}}\, \mbox{\textgamma}_\upmu \bm\psi=q\, \varphi\,  \bm\rho_{\mbox{\tiny ext}}  \mbox{ where }
 \bm\rho_{\mbox{\tiny ext}} =\bm{\psi}^\dagger \bm{\psi}
\end{equation}
Owing to Majorana's condition, Eq.~\ref{mcd}, $\bm{\psi}^\dagger \bm{\psi}=\bm{\psi}^t \bm{\psi}=\bm{\psi}_a \bm{\psi}_a=0$ due to 
anticommutativity. Next let us consider the case of magnetic coupling, namely the hamiltonian density
\begin{equation}
\mathcal{H}=\frac{ m }{4}\, F^{\upmu\upnu}_{\mbox{\tiny ext}} \bar{\bm{\psi}}\, \Sigma_{\upmu\upnu} \bm\psi=- m \, \vec{ B}_{\mbox{\tiny ext}}  \,  \vec{\bm S}  \mbox{ where }
\vec{\bm S}   = \frac{1}{2} \bm{\psi}^\dagger \mbox{\textgamma}^0\, \vec{\Sigma} \bm{\psi}
\end{equation}
Now, using Tab.~\ref{tbl},  it is easy to prove that the following combination of matrices is symmetric
$(\mbox{\textgamma}^0\, \vec{\Sigma})^t= \vec{\Sigma}^t \, (\mbox{\textgamma}^0)^t = \vec{\Sigma} \, \mbox{\textgamma}^0=\mbox{\textgamma}^0\, \vec{\Sigma}$; thus, 
the spin operator $\vec{\bm{S}}$  is zero for a Majorana field, 
again due to anticommutativity of the fermionic fields.

%
%

\subsection{Dirac and Majorana mass in one-neutrino transitions}\label{nan}
We discussed that weak charged interactions, due to their chiral ($V-A$) character,  
distinguish neutrinos and antineutrinos in the limit where their masses are zero, as in the standard model, and 
also in the case in the ultrarelativistic limit, if their masses are not zero as is the case.  
This circumstance in practice limits the possibilities of distinguishing Dirac's neutrinos from Majorana's, as we discuss below.
 
    
Let us begin by considering the  formalism of quantum fields in the case of a Dirac neutrino
\begin{equation}
{\bm{\nu}}_{\mbox{\tiny Dirac}} (t,\vec{x})=\sum_{\vec{p}\lambda} \left[ 
 {\bm{a}}_{\vec{p}\lambda}\ \psi_{\vec{p}\lambda}\! (t,\vec{x})+ 
  {\bm{b}}_{\vec{p}\lambda}^\dagger\  \psi_{\vec{p}\lambda}^*\! (t,\vec{x})
   \right]
\end{equation}
where $\lambda=\pm 1$ is the helicity of the states, $\psi$ are bispinors of plane waves with positive energies:
$\psi_{\vec{p}\lambda}\!(t,\vec{x}) = u_{\vec{p}\lambda}\, e^{ -i (E t- \vec{p}\, \vec{x})/\hbar}$, 
the first term describing the creation of a neutrino in the final state and the second term describing the disappearance of an antineutrino in the initial state.
We have used Majorana's representation of $\mbox{\textgamma}$ matrices, and thus the chirality matrix $\mbox{\textgamma}^5=i \mbox{\textgamma}^0 \mbox{\textgamma}^1 \mbox{\textgamma}^2 \mbox{\textgamma}^3$ 
is also imaginary. When we consider a 
weakly charged interaction, the quantized field is multiplied by the projector $P_L=(1-\mbox{\textgamma}^5)/2$. 
As well-known (see e.g.,~\cite{simon,osc-d}) for ultrarelativistic neutrinos 
the chiral projector selects only bispinors with negative helicity, 
$ P_L\, \psi_{\vec{p}\lambda}   \approx \psi_{\vec{p}\lambda} $ if $\lambda=-1$ and 
$ P_L\, \psi_{\vec{p}\lambda}   \approx  0 $ if $\lambda=+1$.
In correspondence we will have the matrix elements
\begin{equation}  \label{peperon}
\langle \nu_{\vec{p}  \lambda}| 
P_L\,  {\bm{\nu}}_{\mbox{\tiny Dirac}}  (t,\vec{x}) | 0\rangle 
 \approx \left\{ 
\begin{array}{lc}
\psi_{\vec{p}\lambda}  & \mbox{ if }\lambda=-1 \\
 0 & \mbox{ if }\lambda=+1 
\end{array}
\right. \mbox{ when } |\vec{p} |\gg m c
\end{equation}
For antineutrino states we will have $P_L\, \psi_{\vec{p}\lambda}^*=( P_R\, \psi_{\vec{p}\lambda})^*$ and so 
antineutrino states with positive (opposite to neutrino) helicity are selected. 
Neutrinos with positive helicity and antineutrinos with negative helicity do not
interact at all - they are sterile. 

For Majorana fields, the only change is to identify the operators as follows:\footnote{This can also be formally achieved by imposing 
${\bm{\nu}}_{\mbox{\tiny Majorana}}  =( {\bm{\nu}}_{\mbox{\tiny Dirac}}  + {\bm{\nu}}^*_{\mbox{\tiny Dirac}})/\sqrt{2}$,
 thus $    {\bm{c}}= ( {\bm{a}}+ {\bm{b}})/\sqrt{2}$.}
\begin{equation}
{\bm{\nu}}_{\mbox{\tiny Majorana}}  (t,\vec{x})=\sum_{\vec{p}\lambda} \left[ 
 {\bm{c}}_{\vec{p}\lambda}\ \psi_{\vec{p}\lambda}\! (t,\vec{x})+ 
  {\bm{c}}_{\vec{p}\lambda}^\dagger\  \psi_{\vec{p}\lambda}^*\! (t,\vec{x})
   \right]
\end{equation}
The argument  exposed in Eq.~\ref{peperon}
for Dirac field continues to apply, 
with two crucial differences
\begin{itemize}
\item at the order $m/p$, we can have admixtures with `wrong type' particles (antineutrinos rather than neutrinos or vice versa);
\item the operators corresponding to the `sterile' states 
are completely absent.   
\end{itemize}
Therefore, due to the structure of the weak charged interactions,  
in the usual case of ultra-relativistic neutrinos (when the mass of neutrinos is small compared to their 
momentum) the differences between the two types of 
quantized fields disappear.



\subsection{Electron creation and the parameter $\mee$}\label{sth}

Consider the semi-leptonic Hamiltonian density leading to the emission of an electron
$ \mathcal{H}=\sqrt{2} G_{\mbox{\tiny F}}\, \bm{J}^{\mbox{\tiny +}}_\upmu\, \bm{j}^{\mbox{\tiny -}\upmu} $, where the leptonic current is 
\begin{equation}
\bm{j}^{\mbox{\tiny -}}_\upmu= \bar{\bm{e}}\, \mbox{\textgamma}_\upmu \bm{\nu}_{\mbox{\tiny Le}}=
\sum_{j=1}^3 U_{\rm e\,\it j}\ \bar{\bm{e}}\, \mbox{\textgamma}_\upmu  P_{\mbox{\tiny L}}   \bm{\chi}_{j}
\mbox{ with } \bm\chi_j=\bm\chi^*_j
\end{equation}
where we have postulated that the neutrino mass eingestates are Majorana fields. 
The leptonic part of the amplitude, that describes the creation of a couple of electrons, is 
$\langle {\rm e\, e} | {\it T}[    \bm{j}^{\mbox{\tiny -}}_\nu(x) \bm{j}^{\mbox{\tiny -}}_\upmu(y)] | 0\rangle$ and it requires 
to evaluate the contraction $\langle 0  | {\it T}[    \bm{\nu}_{\mbox{\tiny Le}}(x) \, \bm{\nu}_{\mbox{\tiny Le}}^t(y)] | 0\rangle$, namely, an unusual type of 
propagator, that however is non-zero in Majorana's theory. In fact, from 
$\bm{\nu}_{\mbox{\tiny Le}}= U_{\rm e\,\it j}\, P_{\mbox{\tiny L}}   \bm{\chi}_{j}$, used above, and its transpose, written as 
 $\bm{\nu}_{\mbox{\tiny Le}}^t= U_{\rm e\,\it j}\,   \bar{\bm\chi}_j P_{\mbox{\tiny L}}   \mbox{\textgamma}^0$, the  core of the problem reduces to the calculation of 
 an ordinary propagator, namely $\langle 0  | {\it T}[    \bm{\chi}_j (x) \, \overline{\bm{\chi}}_j(y)] | 0\rangle$. The result is 
\begin{equation}
 \langle 0  | {\it T}[    \bm{\nu}_{\mbox{\tiny Le}}(x) \, \bm{\nu}_{\mbox{\tiny Le}}^t(y)] | 0\rangle
  =
    P_{\mbox{\tiny L}}   \mbox{\textgamma}^0
    \int \frac{d^4p}{(2\pi)^4}  \frac{
    i\, U_{\rm e\,\it j}^2\,   m_j \  e^{ -ip(x-y)}}{p^2-m_j^2+i\, 0^+}
 \end{equation}
 The virtual momentum in the denominator has a small time component due to kinematical constraints, 
 whereas the spatial component is of the order of the radius $|\vec{p}|\sim 1/R_0$; therefore, 
 the masses of the light neutrinos  $m_j\ll 100$ MeV are 
 absolutely negligible in the denominator, and the lifetime will depend upon neutrino 
 masses and mixing only through  
 \begin{equation}
\mee=
 \left| \sum_{j=1}^3U_{\rm e\,\it j}^2\,   m_j  \right|  = m_{\upbeta\!\upbeta}
 \end{equation}
 we use the same symbols as the ee-element of the neutrino mass matrix, since the two quantities coincide 
with the phase choice  that makes the ee-element real and non-negative. 
The same quantity is sometimes 
 called `effective neutrino mass' or also `electron neutrino mass',
 most often indicated with the symbol $m_{\upbeta\!\upbeta}$, which recalls the term `(neutrinoless) double beta decay'
 but which does not emphasize the connection with neutrino masses. 
  In this way we have covered the key topic, which started  ideally with \cite{case} in modern times (after $V-A$)
  and which was fully completed in \cite{doi}, with the introduction 
 of $\mee$.  

 To summarise, it is only possible to distinguish between Dirac and Majorana neutrinos if certain mass-specific effects can be observed, i.e., we depend on a parameter whose value is indisputably small.
 
\subsection{A premature dismissal of Majorana's ideas}\label{tooe}

This appendix  reconstructs the story of 
a misunderstanding that began in 1957, the effects of which marred the discussion of Majorana's neutrinos for a long time. 
Let us begin by resuming the present situation. Today we are convinced that   charged weak interactions have a chiral nature, and guided by the principles of the standard model of elementary particles, we think that the only neutrinos needed to discuss these interactions are the left-handed ones. These positions have no implications for the magnitude of the neutrino's Majorana mass, which can be constructed from left-handed neutrinos, and which we intend to explore experimentally. To be sure, Eq.~\ref{conci} consistently describes a non-zero Majorana mass in the context of the (extended) standard model.

In 1957 the discussion was at a much earlier stage, and the 
problem physicists faced was completely different; they had very little idea what the structure of 
weak interactions was, and they needed to proceed. In order to do so, they decided to try and explore 
two hypotheses, i.e., 
{\em 1)~the conservation of the leptonic number and also 2)~the absence of mass of the neutrino,} which were 
consistent with the experimental facts known at the time. 
In this way, a strong and useful  theoretical simplification of the structure of neutrino interactions was achieved, which moreover implied that 
the newly discovered parity violation would apply to neutrinos \cite{pursey,pauli,enz}. These results clarified an important limiting case, 
but of course, they did not help to quantify in any way what the size of the neutrino mass is. 
This point is made very clear in \cite{tous}, a paper of the same year.


Unfortunately, the three seminal articles 
on the chiral structure of neutrino interactions (appeared just previosly) 
included very strong statements, which
left a lasting impression, that masslessness was an inevitable result and not a hypothesis.
E.g.,  in 
\cite{abd} we read: 
{\footnotesize\sc``Parity violation take places for weak-decays in a specified manner which makes the neutrino self-mass 
(like the photon self-mass) vanish''} and 
in \cite{ly2} we read: 
{\footnotesize\sc``A Majorana theory for such a neutrino is therefore impossible.
The mass of the neutrino and the antineutrino
in this theory is necessarily zero''}. 
Even in 
\cite{dau} (that states in the abstract that masslessness is a hypothesis) we read:
{\footnotesize\sc``The mass of the longitudinal neutrino, on the other hand, vanishes automatically''} 
 which is formally correct, as 
`longitudinal neutrino' means just a neutrino
with fixed helicity in modern parlance, 
but presents what is a definition as a fact.

Moreover, we must remember the statement against Majorana's mass made in \cite{mcl} 
(which, as discussed in Eq.~\ref{quibus}, we now consider to be simply wrong) which casts a doubt on the most conventional formalism, that of bispinors. 

As a subsequent examples of prejudices against Majorana, 
we quote from the abstract of an authoritative work of 1959 \cite{prima}: 
{\footnotesize\sc 
``a verdict may be tentatively reached in favour of a `Dirac' neutrino, operationally distinguishable 
from a `Dirac' anti-neutrino, and with conservation of total lepton charge valid in all neutrino interactions''}
and the words of a beautiful paper appeared in 1963 \cite{morita}
{\footnotesize\sc ``Konopinsky 1949 estimated  the transition probabilities for the cases A and B.\footnote{Case A corresponds 
 to Dirac neutrinos, case B to Majorana neutrinos, both with Fermi's interactions.}
 He concluded that the probability  in case B is about thousand times larger
 than that of case A. [...] This contradicts the experimental data of  
double beta decay, which favor the Dirac theory''}.


The problem got solved only slowly, we refer to the papers already cited and that appeared in a variety of contexts \cite{case,whitten,mgs,cuspo,sp,bar,doi,lfl,sv,bk1,bk2}.
However, quite surprisingly, the solution was not appreciated universally  
and some amount of prejudice continued to linger even in recent times.
For example, a justly famous book on nuclear physics \cite{wong}  appeared in 1998 
writes about this: 
{\footnotesize\sc ``one of the interests in double $\beta$-decay [...] 
is to find out whether neutrinos can be Majorana particles. So far all the evidence seems to suggest that they are strictly Dirac particles''},
a statement we do not subscribe for the reasons explained above.

  Interestingly, Majorana's mass always 
  continued to be discussed in the literature {\em but mostly without  bispinors,}  namely  adopting the Weyl's two-dimensional spinor formalism
  as first done in \cite{jehle} (who apparently was not aware of 
Majorana) and above all Case's work \cite{case} (who instead recognizes Majorana). 
Case's paper, which again appeared in 1957, had many merits, among which that of ensuring the consideration/survival of Majorana's ideas in the scientific literature also in the new context: this is why we mention it in Tab.~\ref{bult}. For instance, this work is cited 
 in the classic book~\cite{fy}, 
and also in the first
calculation of nuclear matrix elements for electron creation/double-beta transition of Eq.~\ref{crea}
in the context of  modern $V-A$ theory \cite{whitten}.
Unfortunately, the use of a formalism unfamiliar to a large part of the scientific community in \cite{case} 
created the impression that, 
in order to obtain a coherent field theory, it was {\em necessary} to reformulate Majorana's theory with Weyl spinors. 
However, this is not the case at all: normal bispinors (those with 4 components) are perfectly fine for this purpose, as 
argued in the main text and in this appendix.

A wider and useful discussion of similar issues is given in \cite{silvia}.

\subsection*{Acknowledgments}
I am grateful to Clementina Agodi, Giovanni Benato, Corrado Caselunghe, Silvia de Bianchi, 
Stefano Dell'Oro and Adriano Di Giovanni for useful discussion. 
Work partially supported by the research grant number 2017W4HA7S ``NAT-NET: Neutrino and Astroparticle Theory Network'' under the program PRIN 2017 funded by the Italian Ministero dell'Istruzione, dell'Universit\`a e della Ricerca (MIUR). 
Partly based on presentations \cite{winnete,monnete,siffete}.

  \begin{multicols}{2}

\newpage

%
%
%
%
%
%
%
%
%
%

\end{multicols}

\end{document}